\documentclass[submission,copyright,creativecommons]{eptcs}
\providecommand{\event}{VORTEX 2019}

\usepackage[linesnumbered]{algorithm2e}
\usepackage{amsfonts}
\usepackage{amsmath}
\usepackage{amssymb}
\usepackage[english]{babel} 
\usepackage{breakurl}
\usepackage[inline]{enumitem}
\usepackage{graphicx}
\usepackage{multicol}
\usepackage{subcaption}
\usepackage{tikz}
\usepackage{todonotes}
\usepackage{underscore}
\usepackage{xcolor}
\usepackage{xspace}

\usetikzlibrary{shapes,arrows}
\usetikzlibrary{positioning,automata}
\usetikzlibrary{calc}

\addto\extrasenglish{}
\addto\extrasenglish{}

\newcommand{\keyw}[1]{\textbf{\textsf{#1}}}

\newcommand{\hoaretriple}[3]{\set{#1}\,{#2}\,\set{#3}}
\newcommand{\jmlresult}[0]{\ensuremath{\tt \text{\textbackslash}result}}
\newcommand{\larvatrans}[3]{{#1}\,{\big/}\,{#2}\,{\big/}\,{#3}}
\newcommand{\set}[1]{\{\,{#1}\,\}}

\newcommand{\aspectj}[0]{{\sc AspectJ}\xspace}
\newcommand{\Date}{\textsl{DATE}\xspace}

\newcommand{\elarva}[0]{{\sc Elarva}\xspace}
\newcommand{\key}{Ke\kern-0.1emY\xspace}

\newcommand{\larva}[0]{{\sc Larva}\xspace}

\newcommand{\ppdate}{\textsl{ppDATE}\xspace}

\newcommand{\proactive}[0]{{\sc ProActive}\xspace}
\newcommand{\starvoors}[0]{{\sc StaRVOOrS}\xspace}

\title{Who is to Blame? -- Runtime Verification of \\ Distributed Objects with Active Monitors}

\author{
	Wolfgang Ahrendt
	\institute{Chalmers University of Technology \\ Gothenburg, Sweden}
	\email{ahrendt@chalmers.se}
\and
	Ludovic Henrio
	\institute{Univ Lyon, EnsL, UCBL, CNRS, Inria, LIP \\ Lyon, France}
	\email{ludovic.henrio@ens-lyon.fr}
\and
	Wytse Oortwijn
	\institute{University of Twente \\ Enschede, the Netherlands}
	\email{w.h.m.oortwijn@utwente.nl}
}

\def\titlerunning{Who is to Blame? -- Runtime Verification of Distributed Objects with Active Monitors}
\def\authorrunning{W. Ahrendt, L. Henrio \& W. Oortwijn}

\begin{document}
\maketitle

\begin{abstract}
	Since distributed software systems are ubiquitous,
	their correct functioning is crucially important.
	Static verification is possible in principle, but requires high expertise and effort
	which is not feasible in many eco-systems. Runtime verification can serve as a lean alternative,
	where monitoring mechanisms are automatically generated from property specifications,
	to check compliance at runtime.
	This paper contributes a practical solution for powerful
	and flexible runtime verification of distributed, object-oriented applications,
	via a combination of the runtime verification tool \larva and the active object framework \proactive.
	Even if \larva supports in itself only the generation of local, sequential monitors,
	we empower \larva for distributed monitoring by connecting monitors with active objects,
	turning them into active, communicating monitors.
	We discuss how this allows for a variety of monitoring architectures.
	Further, we show how property specifications, and thereby the generated monitors,
	provide a model that splits the blame between the local object and its environment.
	While \larva itself focuses on monitoring of control-oriented properties,
	we use the \larva front-end \starvoors to also capture data-oriented (pre/post) properties
	in the distributed monitoring.
	We demonstrate this approach to distributed runtime verification with a case study,
	a distributed key/value store.
\end{abstract}

{\section{Introduction}
\label{sec:introduction}

The days of stand-alone software applications are largely over. \emph{Cloud solutions} and \emph{mobile applications} are prominent instances of a general development towards ever more distributed computing. Distributed software is already ubiquitous, and will only grow from here. At the same time, the overwhelming combinatorial complexity of possible interactions and interleavings makes distributed software systems particularly prone to unforeseen, unintended behaviour of multiple criticality. This makes validation efforts even more important than in the stand-alone case. 
Distributed computational scenarios pose enormous challenges to analysis and verification, however. There exist many approaches in the literature, partly supported by tools. But in general, sufficiently powerful static verification approaches tend to be very heavy.

There is a recent trend towards more lightweight formal methods, which are easier to exploit but give limited guarantees. One of them is \emph{runtime verification}, which combines full precision of the execution model (even including the real deployment environment) with full automation. On the other hand, it only ever judges the observed runs, and cannot judge alternative and future runs. Another challenge in runtime verification is the computational overhead of monitoring the running system which can be prohibitive in certain settings.

This paper offers \emph{practical solutions} for flexible runtime verification of distributed, object-oriented applications. Contemporary runtime verification approaches allow users to specify properties on a high level, and hide the details of how the actual monitoring is performed. The active object design pattern allows users to program distributed nodes by writing seemingly sequential code, and hide the details of how the proper communication and coordination between different machines is performed. By combining these two principles, we achieve high-level monitor descriptions and high-level distributed programming at once. Another aim is to allow for a \emph{variety of monitoring architectures} in a natural manner, such that the user can tailor the monitoring architecture to the characteristics of the monitored application and of the underlying network. We achieve this  not only by monitoring active object \emph{applications}, but also by using the active object paradigm \emph{in the monitors themselves}.
Another contribution is the integration of \emph{blame-shifting} into the monitoring, in the spirit of assume-guarantee reasoning. The specification of a node states for every failure whether it is blamed on the monitored node or on its environment. The implementation of the node has to ensure the absence of any failure that is blamed on the node, under the assumption that no failure occurs which is blamed on the node's environment. This supports the localisation of failure while limiting the communication load in the monitoring.
 
Concretely, we made a connection between the runtime verification tool \larva~\cite{CPS09larva} and the active object framework \proactive~\cite{webproactive}. \larva ({\sc L}ogical {\sc A}utomata for {\sc R}untime {\sc V}erification and {\sc A}nalysis) is a tool for monitoring the execution and verifying at runtime the correct behaviour of programs, but it is only adapted to a sequential setting. It could be used to monitor a distributed application but only from a centralised point of view \cite{ColomboFrancalanzaGatt12}, limiting the scalability of the approach. In this work we investigate the usage of \larva to perform \emph{distributed monitoring} of applications. \larva generates a set of monitors for several entities of the observed system and we use \proactive to coordinate the different monitors. \proactive is an active object library that integrates well with \larva because it has no specific syntax for parallelism and distribution: coordination between active objects is performed automatically by the middleware while the programmer only writes standard (sequential) Java code. This way standard \larva monitors can be generated and \proactive  coordinates them in a natural manner.
Further, while \larva itself focuses on monitoring of control-oriented properties, we use the \starvoors ({\sc Sta}tic and {\sc R}untime {\sc V}erification of {\sc O}bject-{\sc Or}iented {\sc S}oftware) front-end for \larva~\cite{Starvoors2015},
to also support the monitoring of data-oriented (pre/post) properties.

The paper is structured as follows. 
\autoref{sec:preliminaries} gives prerequisites on \larva and \proactive. \autoref{sec:approach} discusses the connection of \larva with \proactive to enable distributed monitoring. In particular, our solution allows  \larva monitors to be active objects and enables a variety of monitoring configurations. The notion of blame-shifting and the monitoring of data-oriented properties are also discussed. \autoref{sec:casestudy} applies our approach on a case study: a distributed 
key/value store. 
\autoref{sec:related} compares our work to related research
and \autoref{sec:conclusion} concludes.
}
{\section{Background}
\label{sec:preliminaries}

Rather than writing our tooling from scratch, we combined the \larva runtime verifier with the \proactive programming platform for writing distributed Java applications. This section introduces both \larva and \proactive and outlines some of their aspects that are relevant for the remainder of the paper.

\subsection{Runtime verification with \larva}
\label{sec:larva}

\larva~\cite{CPS09larva} is a runtime verification platform that allows to verify control-flow properties of Java programs, written in an automata-based specification language called \Date ($D$ynamic $A$utomata with $T$imers and $E$vents). \larva generates runtime monitors as Java code out of the automata descriptions of the input properties, and links these monitors to the Java system by using \aspectj.

\begin{figure}[t]
	\centering
	\begin{subfigure}[b]{0.9\textwidth}
		\centering
		\SetInd{0.5em}{0.5em}
		\begin{algorithm}[H]
			\keyw{public class} ${\sf Bank}$ \{ \\ \Indp
				\keyw{public bool} ${\tt login}({\sf String}~code)$ \{ $\cdots$ \} \\
				\keyw{public void} ${\tt logout}({\sf String}~code)$ \{ $\cdots$ \} \\
				\keyw{public void} ${\tt withdraw}({\sf String}~code, \ \keyw{double}~m)$ \{ $\cdots$ \} \\
			\Indm \}
		\end{algorithm}
		\caption{The Java implementation of a small banking system.}
		\label{fig:larva:a}
	\end{subfigure}
	
	\begin{subfigure}[b]{0.9\textwidth}
		\vspace{12pt}
		\centering
		\begin{tikzpicture}[shorten >=1pt, node distance=5.5cm, on grid, initial text={}]
			\node[state,initial] (loggedout) {\small $out$};
			\node[state,accepting, draw=red!50, fill=red!20] (bad) [right of=loggedout] {\small $bad$};
			\node[state] (loggedin) [right of=bad] {\small $in$};
			\path[->] (loggedout) edge [loop above] node {\small $ \larvatrans{\tt login}{\neg \jmlresult}{}$} ();
			\path[->] (loggedout) edge [bend left] node [above] {\small $\larvatrans{\tt login}{\jmlresult}{}$} (loggedin);
			\path[->] (loggedin) edge node [above] {\small $\larvatrans{\tt login}{\sf true}{}$} (bad);
			\path[->] (loggedout) edge [text width=2cm] node [above] {\small $\quad \ \larvatrans{\tt logout}{\sf true}{}$ \\ $\ \larvatrans{\tt withdraw}{\sf true}{}$} (bad);
			\path[->] (loggedin) edge [bend left] node [above] {\small $\larvatrans{\tt logout}{\sf true}{}$} (loggedout);
			\path[->] (loggedin) edge [loop above] node {\small $\larvatrans{\tt get}{\sf true}{}$} ();
		\end{tikzpicture}
		\caption{The \Date property that is used for runtime verification.}
		\label{fig:larva:b}
	\end{subfigure}
	\caption{An example of runtime verification with \larva.}
	\label{fig:larva}
\end{figure}
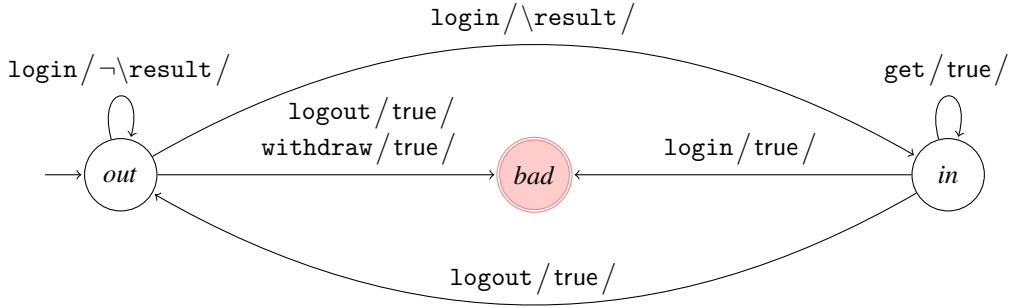

We illustrate the use of \larva with a small example,
a banking system with access control, where users may login, logout, and withdraw money.
\autoref{fig:larva} sketches a \textsf{Bank} class (\ref{fig:larva:a}),
and the \Date property (\ref{fig:larva:b})
that users must first login before being able to logout or withdraw,
and logged-in users should not log in again.

The \Date specification in \autoref{fig:larva:b} consists of three states:~(i) the initial state $out$, describing the logged-out state of the program;~(ii) the normal state $in$ that models the logged-in program state; and~(iii) the bad state $bad$ that models the erroneous state as result of runtime violations. Moreover, \Date transitions are labelled by a triple of the form $\larvatrans{e}{c}{a}$, where: $e$ is an \emph{event} that is connected to method invocations or executions in the program and triggers the transition; $c$ is a \emph{condition} that must hold for the transition to be taken; and $a$ is an \emph{action}, a code snippet that is executed when the transition is taken. In general, the condition- and action components of transitions may access or update global variables.
In fact, since the \larva compiler translates \Date specifications to Java, the $c$ and $a$ components may contain arbitrary Java snippets.

The transitions in \autoref{fig:larva:b} use three different events, namely ${\tt login}$, ${\tt logout}$ and ${\tt withdraw}$, which correspond to invoking the corresponding methods in the implementation. The $\jmlresult$ placeholder in the login transitions is bound to the return value of {\tt login}, so the self-loop in $out$ is taken if the login in the program appeared unsuccessful. The other transitions have ${\sf true}$ as their condition; these transitions are taken unconditionally when triggered. Furthermore, none of the transitions have an associated action, so they are left blank.

\larva automatically generates monitors from \Date property specifications (via \aspectj). We therefore often identify, in the discussion, a \Date property with the monitor generated from it.

\subsection{The \proactive programming platform}
\label{sec:proactive}

\proactive~\cite{webproactive,CH-book} is a Java library for building concurrent and distributed software that implements the \emph{active objects}~\cite{CSUR2017} design pattern. The active objects paradigm---largely inspired by Actors~\cite{Hewitt:1973}---simplifies distributed programming by abstracting concurrency and locality from the programmer. Similar to actors, the primitive unit of computation are objects that have their own thread of control. Active objects may have private state and public methods, and threads may communicate by calling the methods on remote objects. Method invocations are decoupled from method execution, which simplifies the design of distributed systems. By invoking a method, the calling thread pushes a request message into the  queue of the callee thread, which will return a \emph{future} and will eventually process the request by executing the method. This asynchronous construction allows threads to continue working while waiting for remote calls to finish.

The \proactive platform implements the active objects paradigm for Java. In particular, \proactive allows to register Java objects as being ``active'', which exposes the public methods of these objects to other active objects, which may be hosted on different JVMs, possibly located on different machines. Under the hood, when activating a Java object, \proactive spawns a worker thread for the object and constructs a proxy that handles all (network) communication the object gets involved in (method calls on active objects are handled via RMI). However, \proactive hides all these technical details from their users; handling active objects has the same look-and-feel as handling ordinary Java objects.

\begin{figure}[t]
	\begin{algorithm}[H]
${\sf Bank} \ b = ({\sf Bank}){\sf PAActiveObject}.{\tt newActive}({\sf Bank}.class.{\tt getName}(), \, {\tt null})$; \label{line:aoex_newactive} \\
${\sf Account} \ a = b.{\tt createAccount}({\tt userName})$; \\
$b.{\tt deposit}(a,100)$; \\
${\sf Balance}\ m = b.{\tt getBalance}(a)$; \label{line:aoex_rec} \\
${\sf int}\ i = m.{\tt getValue}()$; \label{line:aoex_gv} \\
	\end{algorithm}
	\caption{A simple \proactive example.}
	\label{fig:bankuse}
\end{figure}

\autoref{fig:bankuse} shows a code excerpt illustrating a simple use of \proactive. The first line creates an active object $b$, thus all subsequent uses of object $b$ are active object requests that will be handled asynchronously by the bank object. Each of these call returns a future, stored locally in $a$ and $m$. The code shown in the figure does not use locally the future stored in $a$, but it transmits the future to the bank. The invoker does not need to wait for the future resolution to send it but the bank will probably synchronise on the account creation. It is worth noting that communication between active objects is FIFO, ensuring that all the operations will be handled by the bank in the order of the program. The balance object $m$ receives a future at \autoref{line:aoex_rec}, and the method invocation at \autoref{line:aoex_gv} will be blocked until the balance is obtained (returned by the bank object). 

The support for runtime monitoring is very limited in \proactive. Entry points are provided  to intercept inter-object communications. They are used for example in an execution visualizer, but any precise monitoring of runtime properties has to be done by hand in \proactive (prior to this work).
}
{\section{Monitoring Distributed Objects}
\label{sec:approach}

In the following, we describe how we employ a runtime verification framework for \emph{sequential} applications (\larva) to generate \emph{distributed} monitors for \emph{distributed} (\proactive) objects. We discuss how distributed monitoring is achieved by letting \larva distinguish between method \emph{invocations} and method \emph{executions} on active objects. This section also explains how different monitoring configurations are realised, including orchestrated, centralised and choreography based monitoring, by letting distributed \larva monitors communicate using active objects. Finally, a notion of \emph{blame-shifting} is discussed that is inspired by static assume-guarantee (AG) style reasoning.

\subsubsection*{Running example.}
Throughout this section, we explain and discuss the combination of \larva and \proactive with a distributed version of the banking example discussed in \autoref{sec:larva}, implemented using \proactive. An excerpt of the implementation is given in \autoref{fig:distbank}, where we reuse the \textsf{Bank} class from \autoref{fig:larva}. In fact, this implementation consists of two separate programs, namely:~(i) a server that creates and hosts an active object for the bank, and~(ii) a client that connects to this active object in order to login, logout, or withdraw money.
The client and the server are intended to be instantiated on different JVMs. The only active object, i.e. the only remotely accessible object, is the bank.

\begin{figure}[t]
	\begin{algorithm}[H]
		$\keyw{public class}$ ${\sf Server}$ \{ \\
			\quad \keyw{public static void} ${\tt main}({\sf String}[\,] \ args)$ \{ \\
				\quad\quad ${\sf Bank} \ b = ({\sf Bank}){\sf PAActiveObject}.{\tt newActive}({\sf Bank}.class.{\tt getName}(), \, {\tt null})$; \label{line:ao_newactive} \\
				\quad\quad ${\sf PAActiveObject}.{\tt registerByName}(b, \, args[0])$; \label{line:ao_regactive} \\
			\quad \} \\
		\} \\
		\ \\
		$\keyw{public class}$ ${\sf Client}$ \{ \\
			\quad \keyw{private} \keyw{static} \keyw{void} ${\tt startInterface}({\sf Bank} \ b)$ \{ $\cdots$ \} \\
			\quad \keyw{public static void} ${\tt main}({\sf String}[\,] \ args)$ \{ \\
				\quad\quad ${\sf Bank} \ b = ({\sf Bank}){\sf PAActiveObject}.{\tt lookupActive}($ \label{line:ao_clientconn1} \\
				\qquad\qquad\qquad\qquad ${\sf Bank}.class.{\tt getName}(), \, args[0])$; \label{line:ao_clientconn2} \\
				\quad\quad ${\sf Client}.{\tt startInterface}(b)$; \label{line:ao_clientintf} \\
			\quad \} \\
		\} \\
	\end{algorithm}
	\caption{A distributed implementation of the Banking example, using \proactive.}
	\label{fig:distbank}
\end{figure}

By running the server program,  \proactive  constructs a proxy for the active object in the API-call in \autoref{line:ao_newactive}, and assigns a worker thread to this active object. Any call to methods on $b$ are actually translated to messages sent to this worker, but these details are neatly hidden by \proactive. The API call at \autoref{line:ao_regactive} exposes the new active object $b$ to the outside world, by assigning a network name to it, specified as the argument $args[0]$ in the console. Note that, after executing \autoref{line:ao_regactive}, the thread associated to the active object $b$ keeps on running and is ready to process incoming method invocations.

Another machine might run the ${\sf Client}$ program and connect to the active object hosted by the server (identified by the network name given as an input argument, $args[0]$). The client program will  connect to the specified active object by the API-call in \autoref{line:ao_clientconn1} and \autoref{line:ao_clientconn2}. Observe that the objects $b$ returned in \autoref{line:ao_newactive} and \autoref{line:ao_clientconn1} by \proactive are typed as ${\sf Bank}$ objects: one may use these as ordinary Java objects, even if the object $b$ returned in \autoref{line:ao_clientconn1} is actually  a \emph{stub}---a proxy generated by \proactive that translates all method invocations on $b$ to network messages (actually to RMI communication) to the JVM that physically hosts the bank active object (here the JVM that runs the ${\sf Server}$ program). After connecting to the ${\sf Bank}$ active object, the client may display some user-interface via ${\tt startInterface}$ on \autoref{line:ao_clientintf}, to allow user interaction with the bank.

\subsection{Monitoring distributed objects with \larva}
Even though \larva is not designed for runtime verification of distributed programs, \proactive and \larva make a good match nonetheless, because:
\begin{enumerate}
	\item Although each JVM may host several active objects, each active object only has a single worker thread.
 Therefore, operations on active objects are essentially resolved sequentially. Note that asynchronous communication with futures does not break support for runtime monitoring with \larva.
	\item The \proactive layer hides all details regarding (network) communication between active objects. These details are captured in stub and proxy classes, and those do not break the \aspectj bindings of \larva.
\end{enumerate}

Additionally, many modern distributed programming architectures and protocols can naturally be modelled and constructed with the active objects design pattern, including web and cloud services.

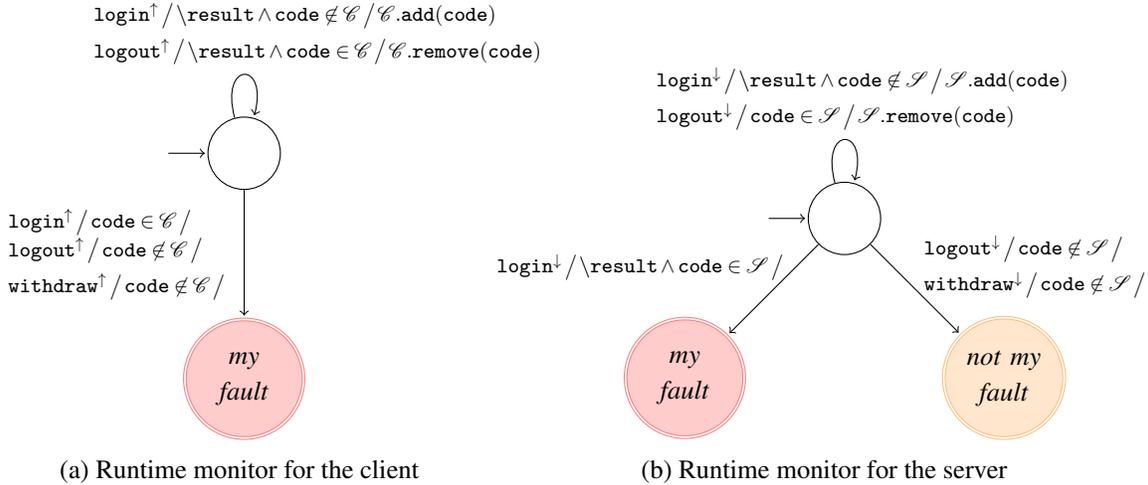
\begin{figure}[t]
	\centering
	\begin{subfigure}[b]{0.40\textwidth}
		\begin{tikzpicture}[shorten >=1pt, node distance=3cm, on grid, initial text ={}]
			\node[state,initial] (working) {};
			\node[state,accepting,draw=red!50,fill=red!20,text width=1cm,align=center] (bad) [below of=working] {\small\emph{my fault}};
	
			\path[->] (working) edge [loop above] node [text width=4cm] {\scriptsize $\larvatrans{{\tt login}^{\uparrow}}{\jmlresult \wedge \texttt{code} \not \in \mathcal{C}}{\mathcal{C}.{\tt add}(\texttt{code})}$ \\ $\larvatrans{{\tt logout}^{\uparrow}}{\jmlresult \wedge \texttt{code} \in \mathcal{C}}{\mathcal{C}.{\tt remove}(\texttt{code})}$} (working);
	
			\path[->] (working) edge node [left, text width=3cm] {\scriptsize $\larvatrans{{\tt login}^{\uparrow}}{\texttt{code} \in \mathcal{C}}{}$ \\ $\larvatrans{{\tt logout}^{\uparrow}}{\texttt{code} \not \in \mathcal{C}}{}$ \\ $\larvatrans{{\tt withdraw}^{\uparrow}}{\texttt{code} \not \in \mathcal{C}}{}$} (bad);
		\end{tikzpicture}
		\caption{Runtime monitor for the client}
		\label{fig:distmon_client}
	\end{subfigure}
	\begin{subfigure}[b]{0.56\textwidth}
		\begin{tikzpicture}[shorten >=1pt, node distance=3cm, on grid, initial text ={}]
			\node[state,accepting,draw=red!50,fill=red!20,text width=1cm,align=center] (bad) {\small\emph{my fault}};
			\node[state,initial] (working) [above right of=bad] {};
			\node[state,accepting,draw=orange!50,fill=orange!20,text width=1cm,align=center] (acc) [below right of=working] {\small\emph{not my fault}};

			\path[->] (working) edge node [left, near start] {\scriptsize $\larvatrans{{\tt login}^{\downarrow}}{\jmlresult \wedge \texttt{code} \in \mathcal{S}}{}$} (bad);
			\path[->] (working) edge [loop above, text width=5cm] node {\scriptsize $\larvatrans{{\tt login}^{\downarrow}}{\jmlresult \wedge \texttt{code} \not \in \mathcal{S}}{\mathcal{S}.{\tt add}(\texttt{code})}$ \\ $\larvatrans{{\tt logout}^{\downarrow}}{\texttt{code} \in \mathcal{S}}{\mathcal{S}.{\tt remove}(\texttt{code})}$} ();
			\path[->] (working) edge node [right, near start, text width=5cm] {\scriptsize \quad  $\larvatrans{{\tt logout}^{\downarrow}}{\texttt{code} \not\in \mathcal{S}}{}$ \\ \quad $\larvatrans{{\tt withdraw}^{\downarrow}}{\texttt{code} \not\in \mathcal{S}}{}$} (acc);
		\end{tikzpicture}
		\caption{Runtime monitor for the server}
		\label{fig:distmon_server}
	\end{subfigure}
	\caption{Runtime monitors for the distributed banking example. Edges with multiple labels abbreviate multiple edges, each with a single label.
}
	\label{fig:distmon}
\end{figure}

\subsubsection{Distributed banking example.}
To monitor the distributed banking implementation in \autoref{fig:distbank} with \larva, separate runtime monitors should be assigned to the client and server program, as they are run on different JVMs. \autoref{fig:distmon} shows the automata-representations of the two \larva monitors.
Both \larva monitors check interaction with the ${\sf Bank}$ active object, but from different perspectives: \autoref{fig:distmon_client} monitors from the client's perspective, whereas \autoref{fig:distmon_server} monitors from the server's perspective. In both cases, the triggers \texttt{login}, \texttt{logout} and \texttt{withdraw} correspond to their eponymous methods in the \textsf{Bank} class.

These runtime monitors again express the property that clients must first login before being able to logout or perform a withdrawal. However, since there are now two different parties involved---a client and a server---we choose, in this example, to monitor the property from two different perspectives.
In the remainder of this section, we discuss the different states and transitions used in \autoref{fig:distmon}, together with their underlying principles.

\paragraph{Call- and execution triggers.}
The \Date specification format allows the distinction between \emph{method calls} and \emph{method executions} when specifying triggers. In \autoref{fig:distmon}, the superscript $\uparrow$ indicates that the transition should be taken upon \emph{calling} the corresponding method in the program, whereas $\downarrow$ indicates an transition triggering upon starting the method \emph{execution}. This distinction between call and execution events is supported by \larva in the sequential setting, but becomes particularly useful when monitoring distributed objects. 

To give an example, when a client invokes a ${\tt login}$ method on a server active object, the client itself does not execute the method  ${\tt login}$. Instead, the invocation becomes a network call to the server. A ${\tt login}^{\downarrow}$ trigger would therefore not be meaningful in the monitor of the client. Moreover, within the server, the ${\tt login}$ method is executed but not invoked. Therefore, a ${\tt login}^{\uparrow}$ trigger is meaningless in the monitor of the server. Generally speaking, a remote call to a method {\tt m} causes two events in different contexts, ${\tt m}^{\uparrow}$ in the context of the caller, and ${\tt m}^{\downarrow}$ in the context of the callee.

To conclude, the call and execution triggers of \Date/\larva match well the events in distributed caller-callee scenarios.

\paragraph{Monitor variables.}
Both runtime monitors in \autoref{fig:distmon} maintain a monitor-local variable, namely a list $\mathcal{C}$ or $\mathcal{S}$ of the codes of logged-in users. Initially these lists are empty. Some transitions update $\mathcal{C}$ or $\mathcal{S}$ via their action component by adding or removing codes. By using state in this way, runtime monitoring can be performed in scenarios where multiple clients log-in multiple users. Moreover, the formal parameter ${\tt code}$ is bound to the matching actual parameter given to the methods of ${\sf Bank}$, and the placeholder $\jmlresult$ captures the return value of the method corresponding to the transition's trigger.

Clarifying the transitions, client monitors go to a bad state when they:
\begin{enumerate*}[label=\textit{(\roman*)}]
	\item perform a login attempt with a code that is already logged-in by the same client, 
          or
	\item perform a logout with a code that is not logged-in, or
	\item attempt to withdraw money on an account that is not logged in.
\end{enumerate*}

The server distinguishes between two kinds of `end' states. The server's monitor goes to the `\emph{my fault}' state when it performs an invalid operation, i.e., upon a successful login for  a user if the monitor knows from its state that the user is already  logged in (i.e., its code is already in $\mathcal{C}$ or $\mathcal{S}$). On the other hand, the monitor goes to the `\emph{not my fault}' state if it detects that a client violates its intended protocol. We use these two different end states for assume-guarantee inspired blame-shifting, see \autoref{fig:blameshifting}.

\subsection{Distributed monitoring configurations}
\label{subsec:configs}

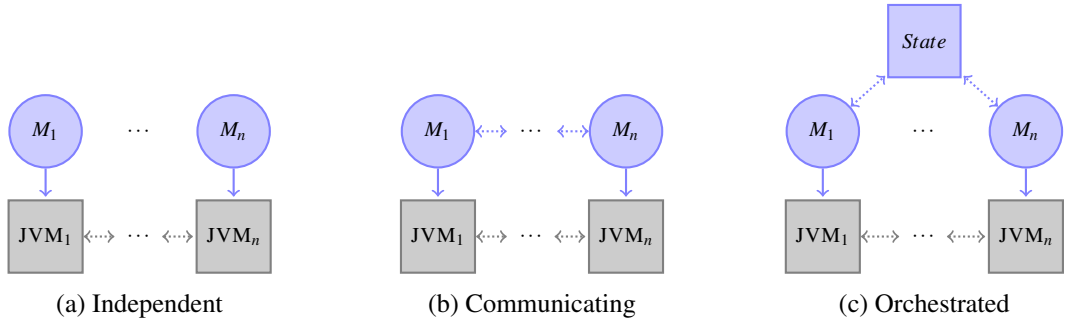
\begin{figure}[t]
	\centering
	\begin{subfigure}[b]{0.32\textwidth}
		\centering
		\begin{tikzpicture}[shorten >=1pt, node distance={1.4cm and 2.5cm}, on grid, initial text ={}]
			\node[state,rectangle,draw=black!50,fill=black!20,thick] (jvm1) {\scriptsize\sc $\text{JVM}_1$};
			\node[state,rectangle,draw=black!50,fill=black!20,thick] (jvm2) [right = of jvm1] {\scriptsize\sc $\text{JVM}_n$};
			\node (jvmdots) at ($(jvm1)!0.5!(jvm2)$) {\scriptsize $\cdots$};
			
			\node[state,draw=blue!50,fill=blue!20,thick] (mon1) [above =of jvm1] {\scriptsize $M_1$};
			\node[state,draw=blue!50,fill=blue!20,thick] (mon2) [above =of jvm2] {\scriptsize $M_n$};
			\node (mondots) at ($(mon1)!0.5!(mon2)$) {\scriptsize $\cdots$};
	
			\path[<->,thick,draw=black!50,densely dotted] (jvm1) edge node {} (jvmdots);
			\path[<->,thick,draw=black!50,densely dotted] (jvmdots) edge node {} (jvm2);
			\path[->,thick,draw=blue!50] (mon1) edge node {} (jvm1);
			\path[->,thick,draw=blue!50] (mon2) edge node {} (jvm2);
		\end{tikzpicture}
		\caption{Independent}
		\label{fig:distmon1}
	\end{subfigure}
	\begin{subfigure}[b]{0.32\textwidth}
		\centering
		\begin{tikzpicture}[shorten >=1pt, node distance={1.4cm and 2.5cm}, on grid, initial text ={}]
			\node[state,rectangle,draw=black!50,fill=black!20,thick] (jvm1) {\scriptsize\sc $\text{JVM}_1$};
			\node[state,rectangle,draw=black!50,fill=black!20,thick] (jvm2) [right = of jvm1] {\scriptsize\sc $\text{JVM}_n$};
			\node (jvmdots) at ($(jvm1)!0.5!(jvm2)$) {\scriptsize $\cdots$};
			
			\node[state,draw=blue!50,fill=blue!20,thick] (mon1) [above = of jvm1] {\scriptsize $M_1$};
			\node[state,draw=blue!50,fill=blue!20,thick] (mon2) [above = of jvm2] {\scriptsize $M_n$};
			\node (mondots) at ($(mon1)!0.5!(mon2)$) {\scriptsize $\cdots$};
	
			\path[<->,thick,draw=black!50,densely dotted] (jvm1) edge node {} (jvmdots);
			\path[<->,thick,draw=black!50,densely dotted] (jvmdots) edge node {} (jvm2);
			\path[->,thick,draw=blue!50] (mon1) edge node {} (jvm1);
			\path[->,thick,draw=blue!50] (mon2) edge node {} (jvm2);
			\path[<->,thick,draw=blue!50,densely dotted] (mon1) edge node {} (mondots);
			\path[<->,thick,draw=blue!50,densely dotted] (mon2) edge node {} (mondots);
		\end{tikzpicture}
		\caption{Communicating}
		\label{fig:distmon2}
	\end{subfigure}
	\begin{subfigure}[b]{0.32\textwidth}
		\centering
		\begin{tikzpicture}[shorten >=1pt, node distance={1.4cm and 2.7cm}, on grid, initial text ={}]
			\node[state,rectangle,draw=black!50,fill=black!20,thick] (jvm1) {\scriptsize\sc $\text{JVM}_1$};
			\node[state,rectangle,draw=black!50,fill=black!20,thick] (jvm2) [right = of jvm1] {\scriptsize\sc $\text{JVM}_n$};
			\node (jvmdots) at ($(jvm1)!0.5!(jvm2)$) {\scriptsize $\cdots$};
			
			\node[state,draw=blue!50,fill=blue!20,thick] (mon1) [above = of jvm1] {\scriptsize $M_1$};
			\node[state,draw=blue!50,fill=blue!20,thick] (mon2) [above = of jvm2] {\scriptsize $M_n$};
			\node (mondots)  at ($(mon1)!0.5!(mon2)$) {\scriptsize $\cdots$};
			\node[state,rectangle,draw=blue!50,fill=blue!20,thick] (monstate) [above = 1.2cm of jvm2] at ($(mon1)!0.5!(mon2)$) {\scriptsize $State$};
	
			\path[<->,thick,draw=black!50,densely dotted] (jvm1) edge node {} (jvmdots);
			\path[<->,thick,draw=black!50,densely dotted] (jvmdots) edge node {} (jvm2);
			\path[->,thick,draw=blue!50] (mon1) edge node {} (jvm1);
			\path[->,thick,draw=blue!50] (mon2) edge node {} (jvm2);
			\path[<->,thick,draw=blue!50,densely dotted] (mon1) edge node {} (monstate);
			\path[<->,thick,draw=blue!50,densely dotted] (mon2) edge node {} (monstate);
		\end{tikzpicture}
		\caption{Orchestrated}
		\label{fig:distmon3}
	\end{subfigure}
	\caption{Different monitoring configurations:~\ref{fig:distmon1} depicts purely independent monitors,~\ref{fig:distmon2} shows monitoring with communication, and~\ref{fig:distmon3} depicts orchestrated monitoring, where some centralised component $State$ maintains the global state.}
	\label{fig:distmons}
\end{figure}

The runtime monitoring setup that we considered so far is purely distributed, as illustrated in \autoref{fig:distmon1}. Each participating JVM  ($\text{JVM}_i$) has its own \larva monitor ($\text{M}_i$), and these monitors do not communicate with each other. However, there are many more monitoring configurations proposed in the literature~\cite{FRANCALANZA2013186,ColomboF2016}, some of which rely on monitor communication. 
Indeed,  if independent monitors are by nature extremely efficient, inter-monitor communication is sometimes necessary to establish correctness. We explain below how to enable monitor interactions in our context and what  interaction patterns we envision for monitors.


\paragraph{Distributed monitor communication.}
Recall that \larva monitors are translated to Java code (to be bound to the monitored implementation using \aspectj), and that the transition in the automata may contain arbitrary snippets of Java. We exploit this to realise monitor communication, as shown in \autoref{fig:distmon2}, by instantiating active objects \emph{inside the \larva monitors} and have them connect to each other. Runtime monitors may call the public methods on the active objects of remote monitors and thereby influence their state. Moreover, since active objects are  used as ordinary Java objects, the \larva monitors may define triggers on the public methods of their active objects. By doing this, a transition can be taken when a remote monitor invokes a public method on a local active object, thus implementing monitor synchronisation.


\begin{figure}[t]
	\centering
	\begin{subfigure}[b]{0.45\textwidth}
		\centering
		\begin{tikzpicture}[shorten >=1pt, node distance={3cm and 2.5cm}, on grid, initial text={}]
			\node[state] (working) {};
			\node[state,accepting,draw=red!50,fill=red!20] (bad) [right of=working] {\scriptsize $bad$};
			\node[state,draw=none,xshift=0.75cm,yshift=-2cm] (aux1) [above left of=working] {};
			\node[state,draw=none,xshift=0.75cm,yshift=2cm] (aux2) [below left of=working] {};
			\path[->,densely dashed,color=gray] (aux1) edge [bend left] node {} (working);
			\path[->,densely dashed,color=gray] (aux2) edge [bend right] node {} (working);
			\path[->] (working) edge [bend left] node [above] {\scriptsize $\larvatrans{{\color{gray}-}}{{\color{gray}-}}{r.{\tt notify}()}$} (bad);
		\end{tikzpicture}
		\caption{Client monitor snippet}
		\label{fig:monitorcomm1}
	\end{subfigure}
	\begin{subfigure}[b]{0.45\textwidth}
		\centering
		\begin{tikzpicture}[shorten >=1pt, node distance={3cm and 2.5cm}, on grid, initial text={}]
			\node[state] (working) {};
			\node[state] (bad) [right of=working] {};
			\node[state,draw=none,xshift=0.7cm,yshift=2.15cm] (aux1) [below left of=working] {};
			\node[state,draw=none,xshift=-0.8cm,yshift=2.4cm] (aux2) [below right of=bad] {};
			\node[state,draw=none,xshift=-0.8cm,yshift=-2cm] (aux3) [above right of=bad] {};
			\path[->,densely dashed,color=gray] (aux1) edge [bend right] node {} (working);
			\path[->,densely dashed,color=gray] (bad) edge [bend right] node {} (aux2);
			\path[->,densely dashed,color=gray] (bad) edge [bend left] node {} (aux3);
			\path[->] (working) edge [bend left] node [above] {\scriptsize $\larvatrans{{\tt notify}^{\downarrow}}{{\color{gray}-}}{{\color{gray}-}}$} (bad);
		\end{tikzpicture}
		\caption{Server monitor snippet}
		\label{fig:monitorcomm2}
	\end{subfigure}
	\caption{Synchronisation of monitors using active objects. Here $r$ is an active object with a public method $\texttt{notify}()$, that is instantiated by the server monitor and called by the client monitor (for notifying the server monitor).}
	\label{fig:monitorcomm}
\end{figure}
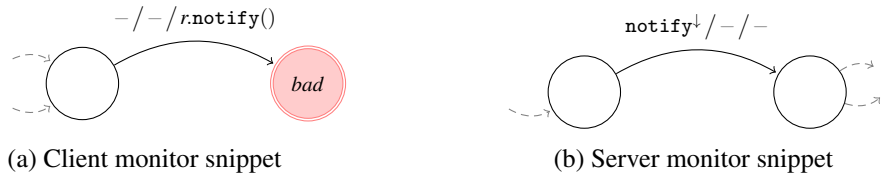

\autoref{fig:monitorcomm} illustrates the integration of synchronisation between independent distributed \larva monitors using \proactive (this illustration might for example extend the setup from \autoref{fig:distmon}). The server monitor  may \emph{itself} instantiate and host a new active object $r$ (with a public method {\tt notify}), and the client monitor may connect to this active object, like illustrated in \autoref{fig:distbank}. Then when the client monitor transitions to some bad state for example, it may call $r.{\tt notify}()$ in the action component of the transition (\ref{fig:monitorcomm1}). The object $r$ in the client monitor is only a proxy object generated by \proactive, so this method invocation targets the monitor of the server. Since {\tt notify} is an ordinary Java method, the server's monitor may have a trigger for its execution (i.e. ${\tt notify}^{\downarrow}$), so that the monitor can take a transition when the client invokes {\tt notify} (\ref{fig:monitorcomm2}).

\paragraph{Orchestrated monitoring.}
By allowing monitors to communicate with each other, we can also realise  elaborate monitoring configurations like \emph{orchestrated monitoring}, illustrated in \autoref{fig:distmon3}. Instead of having the monitors communicate with each other directly, they communicate with a \emph{centralised} active object (called \emph{State} in the figure). This schema is the best way to implement a globally shared memory. Indeed active objects do not share data, this organisation can allow us to an active object to store a global state, together with getters and setters invokable by the distributed monitors. In addition, since the centralised component is an active object, and therefore plain Java, it may have its own \larva monitor and verify invariants on the global state. 

These different configurations open-up a design space of distributed monitoring, as they may be combined in many ways. For example, one may consider orchestrated monitoring as in~\ref{fig:distmon3}, but still allow \larva monitors to communicate with each other, as in~\ref{fig:distmon2}. Moreover, if a single orchestrator (like shown in~\ref{fig:distmon3}) does not provide enough scalability, one may consider a hierarchical structure of orchestrators, each coordinating a subset of the distributed monitors.

\subsection{Assumption-guarantee style blame-shifting}
\label{fig:blameshifting}

In some scenarios, independent distributed monitoring is insufficient to establish global correctness, while monitor communication is performance-wise too heavyweight. The distributed hash table case study in \autoref{sec:casestudy} falls into this category, since communication-based monitoring would break its scalability, while independent monitoring alone is not enough to establish correctness globally.

For these scenarios we propose a notion of \emph{blame-shifting} that is inspired by assume-guarantee (AG) reasoning~\cite{335c34f7a1d744fa8981c64704e889d6}. The idea is that, instead of having runtime monitors rely on network communication, they could instead rely on \emph{assumptions} from the environment, while giving certain \emph{commitments} to the environment in return.
This system of assumptions and commitments must be consistent: a monitor may only assume properties from the environment that are guaranteed/committed by the monitors of the environment. (Otherwise, the insights from the runtime verification would be limited.)

This implementation of AG reasoning constitutes a `my fault/not my fault' blame system. A `not my fault' occurs when a monitor observes
interaction with remote objects that 
violates its assumptions on the environment.
A `my fault' occurs when a monitor either observes
interaction with a remote object that violate its commitment to the environment, or if it violates a local sub-property (possibly leading to interaction that violates its commitment to the environment).
Moreover, if the assumptions and commitments of interacting distributed objects are in sync, then a `not my fault' in some objects will be mirrored by a `my fault' in one or several other objects.
In conclusion, AG style monitoring can be used to detect and locate distributed runtime violations, without having to resort to centralised monitoring or other configurations that require communication.

\paragraph{Blame-shifting by example.}
An example of blame-shifting was already presented in \autoref{fig:distmon}. Here, the server monitor distinguishes between violations from the environment and local violations via two different end states, named `\emph{not my fault}' and `\emph{my fault}'. The client monitor does not have a `\emph{not my fault}' state in this example, as the intended protocol only really restricts the client (e.g. the \emph{client} must not withdraw before logging in). Also notice that the assumptions of the server are consistent with the commitments of the client; if the server's monitor goes to the `\emph{not my fault}' state, it means that a client performed a logout or a withdrawal without being logged-in, and therefore goes to its `\emph{my fault}' state.

Although the two runtime monitors described in \autoref{fig:distmon} do not communicate, the server can still determine whether clients violated their protocol. This concept of blame-shifting is demonstrated further in the case study (\autoref{sec:casestudy}) in particular in a system consisting of more than two objects.


\paragraph{Global consistency.}
At the moment we do not have a mechanised way of checking whether the assumptions made by \larva monitors are consistent with the commitments of the monitors of the environment. This currently has to be established manually. We are investigating ways to make this check mechanical, as a static verification task. This would result a combination of \emph{static global} consistency checking and \emph{local runtime} verification.

\subsection{Monitoring data-oriented aspects}
\label{sec:hoaretriples}
Distributed systems typically consist of nodes that perform local computations and distribute the computed results over the network, via protocols. Runtime verification of distributed systems should therefore cover both the computation and the distribution of data. Even though \larva allows \Date specifications to use data to a limited extend, it is mostly focused on control-oriented properties.

In order to extend our approach to the monitoring of data-oriented properties, we define \ppdate, an extension of \Date with \emph{Hoare triples} (\emph{pp} stands for \emph{p}re/\emph{p}ost-condition), as supported by the 
\starvoors~\cite{AhrendtChimentoPaceSchneider17a,Starvoors2015} tool ({\sc Sta}tic and {\sc R}untime {\sc V}erification of {\sc O}bject-{\sc Or}iented {\sc S}oftware).
Hoare triples are of the standard form $\hoaretriple{P}{{\tt m}}{Q}$, with $P$ and $Q$ assertions expressed in first-order logic and ${\tt m}$ the header of a Java method.
In \ppdate, Hoare-triples are \emph{state dependent}: each state in the runtime monitor automaton contains its own set of Hoare triples that have to hold in that state. \starvoors translates \ppdate to \Date and uses \larva thereafter.\footnote{\starvoors also supports monitor optimisation through (partial) static verification results, a feature which we do not use yet in the distributed setting.}
\autoref{sec:casestudy} illustrates  how Hoare triples can make the specification and monitoring of distributed applications easier.}
{\section{Case Study: ActiveCAN}
\label{sec:casestudy}

This section demonstrates our verification approach on a case study: verifying the correctness of a CAN~\cite{Ratnasamy01:CAN}---a Content-Addressable Network---implemented with active objects
in the \proactive platform~\cite{PHBA13CAN}.
In the sequel we refer to this case study as ActiveCAN.
The main motivations for considering the ActiveCAN case study are:~(i) the case study is external, rather than a toy example that we constructed ourselves (an evolved version has also been used in a large-scale application)\footnote{See \url{http://play.ow2.org}.};~(ii) the property to be runtime verified is a combination of both data- and control-oriented properties; and~(iii) by nature the example consists in many identical peers, which highlights the meaning of the distinction 'my fault' vs. 'not my fault', where 'not my fault' means either another peer or the rest of the system. 

\paragraph{Application description.}
A CAN is a distributed infrastructure that provides hash table-like functionality. The basic operations of a CAN are: ${\tt insert}(k,v)$ for inserting a value $v$ at key $k$, and ${\tt lookup}(k)$ for looking-up the possible value stored at key $k$. The originality is that keys are tuples. CANs consist of many  peers that are connected via a network, and each peer owns a fragment of the entire key space, i.e. a hyperrectangle. When a new peer joins the network, the key space in the CAN is split, allowing the joined peer to receive ownership of the new fragment. Moreover, when performing an {\tt insert} or {\tt lookup} on a peer, the operation may either be handled locally if the key is owned by the local peer, or otherwise be relayed to a remote peer. For relaying operations to remote peers the CAN infrastructure contains a routing protocol that is  scalable.

The ActiveCAN consists of a set of active objects, each responsible for the storage of data indexed in one hyperrectangle. Each peer knows the zone it manages, but also the zones of the neighbouring peers to route messages to the right target.
The ActiveCAN peers provide three operations:
\begin{itemize}
	\item ${\tt join}(p)$, to add one new peer $p$ in the network. The joined peer will split its hyperrectangle into two parts and delegate one to the new peer.
	\item ${\tt insert}(k,v)$, to insert a new value $v$ at a given key $k$ where $k$ is a tuple.
	\item ${\tt lookup}(k)$, to fetch the value previously stored at a given tuple $k$.
\end{itemize}

Communication between neighbouring peers is performed by method calls on active objects, and each request is transmitted this way to the adequate target. Finally, lookup relies on futures to return the fetched value. 

\paragraph{Verified properties.} 
Essentially, the following high-level properties are addressed:
\begin{enumerate}
	\item The behaviour of the hash table is \emph{consistent}, meaning that:
		\begin{enumerate*}[label=\textit{(\roman*)}]
			\item after calling ${\tt insert}(k,v)$ the data element $v$ is stored somewhere in the network at key $k$,
			\item if ${\tt lookup}(k)$ returns a positive result, then $k$ is mapped somewhere in the network,
			\item if ${\tt lookup}(k)$ returns a negative result, then $k$ is not mapped in the network.
		\end{enumerate*} \label{itm:case_prop1}
	\item There are no cycles in the routing protocol. In particular, we verify that:
		\begin{enumerate*}[label=\textit{(\roman*)}]
			\item any ${\tt lookup}$ and ${\tt insert}$ operation is either handled locally by a peer, or the peer has at least one neighbour to defer the operation to, and
			\item if a ${\tt lookup}$ or ${\tt insert}$ needs to be remotely resolved, the operation is deferred to a remote peer that is \emph{closer} to the target peer (hence there are no cycles in the routing protocol).
		\end{enumerate*} \label{itm:case_prop2}
\end{enumerate}

\medskip
Property \eqref{itm:case_prop1} in the above is data-oriented, whereas \eqref{itm:case_prop2} is control-oriented, and both are necessary to establish correctness of the overall infrastructure. 
The case study thereby highlights the necessity for verification techniques to include data and control aspects and motivates the usage of the \starvoors front-end to \larva. The distributed setting is particularly interesting here because peers typically perform some computation over data and distribute requests and results via some routing protocol.
\larva generates monitors checking these properties, reporting when any of them is violated while using the hash table. 

%
%

%

To verify property \eqref{itm:case_prop1}, for each peer, the runtime monitor  maintains a list $\mathcal{K}$ of keys to be mapped in the key/value store. For each key, it also maintains a boolean flag  that indicates whether the key is stored locally or remotely. Therefore, $\mathcal{K}$ contains all keys stored locally by the peer, but also records the keys of remote ${\tt lookup}$ and ${\tt insert}$ operations that routed through the peer.

The runtime monitors use blame-shifting in the style of AG reasoning to determine whether a lookup or insert is handled correctly. In more detail, when a locally resolved ${\tt lookup}$ behaves incorrectly, the runtime monitor of the local peer is able to detect this: it is classified as `my fault'. When a remotely resolved ${\tt lookup}$ behaves incorrectly (e.g. stating the key is unmapped while the runtime monitor knows from $\mathcal{K}$ that the key has already been stored), then the local peer can determine that another peer has violated the property: it is classified as `not my fault'. Since the remotely resolved ${\tt lookup}$ has been deferred to a neighbouring peer, that peer is closer to figuring out which peer actually violated the property. In particular, due to the consistency of the AG-style assumptions and commitments, the neighbouring peer's monitor must also be in a \emph{not my fault} or \emph{my fault} state. This constitutes a chain of blames, ending-up in the misbehaving peer (i.e. the peer that is in the \emph{my fault} state).

Property \eqref{itm:case_prop2} is verified using the Hoare triple extensions that are discussed in \autoref{sec:hoaretriples}. Every time a ${\tt lookup}$ or ${\tt insert}$ operation needs to be deferred to a remote node, the implementation calls the helper method ${\tt getZoneClosestTo}(k)$, which returns the neighbouring zone that is closest to the destination peer. The property of loop freedom in the routing protocol is captured by the following Hoare triple.
\begin{equation*}
	\big\{ {\sf true} \big\} \ {\tt getZoneClosestTo}(k) \ \bigg\{ \begin{array}{c} \backslash{result} \not = {\sf null}  \ \wedge \\  \backslash{result}.{\tt distance}(k) < localzone.{\tt distance}(k) \end{array} \bigg\}
\end{equation*}
Every zone $z$ has a distance function $z.{\tt distance}(k)$ that calculates the Euclidean distance between the center of $z$ and the specified key $k$. The variable $localzone$ refers to the zone of the callee peer.
This Hoare triple is checked every time the ${\tt getZoneClosestTo}$ method is called in the implementation. This method returns an object of type ${\sf Zone}$---the zone of the neighbour closest to the destination peer. 

\paragraph{Availability.}
The source code of the ActiveCAN case study, together with verification instructions, are available online\footnote{\url{https://github.com/utwente-fmt/RV2018-ActiveCAN}.}. Our experiments did not reveal runtime violation of properties \eqref{itm:case_prop1} and \eqref{itm:case_prop2}.}
{\section{Related Work}
\label{sec:related}


Even if the area of runtime verification is much more elaborate in the stand-alone setting\footnote{This includes the analysis of single nodes, only, running in a distributed environment.}, runtime verification of distributed systems is a very active area of research. The recent article~\cite{Francalanza-RV2018} provides an excellent overview over scenarios and characteristics of this sub-field, and discusses the existing approaches on that background. Here, we only include works with sufficient overlap in aim, method, or application area. Generally, our discussion of configurations in \autoref{subsec:configs} is consistent with the monitor organisation categories in~\cite{Francalanza-RV2018}. However, compared to the analysis there the \larva methodology for specification by nature splits the monitoring into properties specified and verified for each object, which naturally extends to a truly distributed monitoring approach.
A recent work use distributed monitors to control the executions of Erlang actors~\cite{FMT-Dais18}, the behaviour of the program is expressed as a choreography and the monitors control the execution so that, in case of failure, the application can rollback to a safe state and then run along a different path. 
Compared to these choreography-based approaches, we provide the monitoring of a reactive system that does not have a service choreography specification. For example we are able to monitor a peer-to-peer system that has a more concurrent behaviour than choreographies.
In such a system it makes more sense to specify the correctness of each entity based on assumptions on the environment than from a global perspective, and this naturally provides distributed monitors. However, it is more difficult to state properties of the global   system in our approach than from a choreographed perspective, but to the best of our knowledge, no other runtime-verification approach is able to verify at runtime a reactive system made of distributed objects.


The \elarva runtime verifier~\cite{ColomboFrancalanzaGatt12} is an adaption of \larva for monitoring concurrent Erlang programs. In particular, \elarva adapts \larva by translating the object-oriented monitoring constructs to a process-oriented setting, and gives an asynchronous interpretation to \larva's monitoring semantics. However, \elarva only supports centralised monitoring and is unable to monitor across multiple machines. Our implementation allows both centralised and decentralised monitoring, and allows distributed monitors to communicate and coordinate using active objects, as described in \autoref{sec:approach}. Another notable difference is that we did not change the synchronous \larva's monitoring semantics, and instead leave it to \proactive to `hide' the asynchronous nature of the underlying communication (not only from the programmer but also from \larva).



There are several tools for static verification of active objects, relying on model-checking~\cite{AmeurBoulifa2017,DBLP:journals/fuin/SirjaniMSB04}, static analysis~\cite{AlbertAFGGMPR14}, behavioural types~\cite{HLM-IFM2017}, or deductive techniques~\cite{DinTHJ15}, but the support for monitoring and runtime verification of active object systems is quite weak. Basic monitoring tools exist for actors and active object systems. For example Akka~\cite{AkkaBook} provides an interesting hierarchy of actors for monitoring failures, both ABS and \proactive feature a tool for viewing or debugging active object execution (see e.g. Section~3.3 of~\cite{HR-lmcs17}), and a monitoring framework has been developed for Erlang~\cite{CF-iFM16}. 
In existing platforms runtime verification of functional correctness for active object applications must be done by hand. We believe that we can fit our approach to the infrastructures proposed in actor monitoring systems by generating monitors specific to an actor monitoring framework, like~\cite{CF-iFM16}; however our current approach has the crucial advantage to minimise the changes to the monitor generation of \larva.

The choice of locations of the monitors is quite an important
issue because communication across locations is usually expensive and information-sensitive. A good discussion about this choice is presented in~\cite{FRANCALANZA2013186}, where a theoretical framework is presented for comparing those choices. Such a discussion is complementary to the solution presented in this system paper, which shows how different monitoring architectures can be naturally realised for a distributed variant of Java.

We are not aware of other work which adapts the assumption-guarantee paradigm, known from compositional static verification, to the runtime verification setting.}
{\section{Conclusion}
\label{sec:conclusion}

This article shows the importance of two key challenges in monitoring of distributed applications: distributed monitoring by independent active monitors, and verification of properties mixing data-oriented and control-oriented aspects.

From a technical point of view, this article provides two contributions addressing these two challenges. First it presents an effective distributed monitoring mechanism. This is realised thanks to active monitors that are generated by our framework. Our runtime verification environment combines monitor generation of the \larva  framework with an active object middleware, \proactive. This way, the standard Java code generated by \larva generates distributed active monitors at runtime to detect the violation of safety properties in a distributed monitoring fashion. The verification setup integrates assumption-guarantee-style blame shifting to efficiently localise runtime failures while limiting communication between monitors. Moreover, the Hoare triple extension to \larva provides to the programmer a better abstraction for specifying properties mixing data-oriented and control-oriented aspects. This is particularly relevant in connection with the active object paradigm, where a task is the execution of a method.


The approach has been illustrated by monitoring a distributed peer-to-peer system implementing a key-value store. This example and the properties we were able to monitor illustrate well the contributions of this work: it is by nature distributed, and needs a distributed monitoring infrastructure; and the application mixes control-oriented and data-oriented properties both in the routing of messages and in the storage/retrieval of data; and the blame-shifting allows to trace the cause of an error along the routing path (even if this is error tracing is currently done on the meta-level and not yet automated).

Our approach is by nature distributed. In particular, we do not favour verification of global properties and prefer to focus on a local rely-guarantee approach. The counterpart is that our framework is not particularly adapted to reason at a global level, and in particular an object is unable to distinguish if an external error is due to another object or to a communication error. However, as monitors can communicate and  a  global state can be stored, we should be able in the future to extend our framework with a better support for global properties and better classification of external errors.

In the future, we want to investigate the runtime overhead of our distributed monitoring methodology and its variants. Also, we want to explore how easy/difficult it is for users to specify the properties of interest in the suggested style. Another line of future work is to migrate the \starvoors approach \cite{AhrendtChimentoPaceSchneider17a} to the distributed setting, to combine static verification with our distributed runtime verification method. Just like in \starvoors, this has great potential to decrease the runtime overhead and increase the scalability.

}

\subsubsection*{Acknowledgements.}
The authors would like to thank Gordon Pace and Gerardo Schneider
for fruitful discussions in the course of this work,
and Mauricio Chimento for implementing some adaptions in the \starvoors tool.
This work is partially supported by the NWO TOP 612.001.403 project VerDi,
and by the COST Action IC1402 Runtime Verification beyond Monitoring.

\bibliographystyle{eptcs}
\bibliography{../common/vercors,references}

\end{document}